\documentclass[submission, Phys]{SciPost}

\hypersetup{
    colorlinks,
    linkcolor={red!50!black},
    citecolor={blue!50!black},
    urlcolor={blue!80!black}
}

\usepackage[bitstream-charter]{mathdesign}
\usepackage{bm}
\usepackage{empheq}

\DeclareSymbolFont{usualmathcal}{OMS}{cmsy}{m}{n}
\DeclareSymbolFontAlphabet{\mathcal}{usualmathcal}

\newcommand{\be}{\begin{equation}}
\newcommand{\ee}{\end{equation}}
\newcommand{\bea}{\begin{eqnarray}}
\newcommand{\eea}{\end{eqnarray}}
\newcommand{\w}{\omega}

\newcommand{\Tr}{{\rm Tr}}
\renewcommand{\Im}{{\rm Im}}
\renewcommand{\Re}{{\rm Re}}

\newcommand{\p}{{\prime}}
\newcommand{\pp}{{\prime\prime}}
\newcommand{\ppp}{{\prime\prime\prime}}

\newcommand{\up}{\uparrow}
\newcommand{\dn}{\downarrow}

\newcommand{\Hxc}{{\rm Hxc}}
\newcommand{\xc}{{\rm xc}}
\newcommand{\It}{\tilde{I}}
\newcommand{\sgn}{{\rm sgn}}

\newcommand{\aK}{a_{\rm K}}

\begin{document}

\begin{center}{\Large \textbf{\color{scipostdeepblue}{
Doping-driven Mott transition from steady-state density functional theory 
\\
}}}\end{center}

\begin{center}\textbf{
David Jacob\textsuperscript{1$\star$} and
Stefan Kurth\textsuperscript{2,3,4}
}\end{center}

\begin{center}
  {\bf 1} Departamento de F\'{i}sica, Universidad de Alicante, Campus de San
  Vicente del Raspeig, E-03690, Alicante, Spain
\\
{\bf 2} Nano-Bio Spectroscopy Group and European Theoretical Spectroscopy
  Facility (ETSF), Departamento de Pol\'imeros y Materiales Avanzados:
  F\'isica, Qu\'imica y Tecnolog\'ia, Universidad del Pa\'is Vasco EHU,
  Avenida de Tolosa 72, E-20018 San Sebasti\'an, Spain
\\
{\bf 3} IKERBASQUE, Basque Foundation for Science, Plaza Euskadi 5,
  E-48009 Bilbao, Spain
\\
{\bf 4} Donostia International Physics Center (DIPC), Paseo Manuel de
  Lardizabal 4, E-20018 San Sebasti\'{a}n, Spain 
\\[\baselineskip]
$\star$ \href{mailto:email1}{\small david.jacob@ua.es}
\end{center}

\section*{\color{scipostdeepblue}{Abstract}}
\textbf{\boldmath{%
We describe the doping-driven Mott transition in the Hubbard model
within the framework of steady-state density functional theory, or i-DFT. 
In order to access the many-body spectral function in i-DFT, an
approximation to the exchange-correlation (xc) bias at arbitrary density and
current is required. 
By making use of Fermi-liquid theory, we derive conditions on the xc bias of 
i-DFT in terms of the quasiparticle weight, thus establishing a clear connection 
between i-DFT and Fermi-liquid theory.
The Fermi liquid conditions are then employed to guide the construction
of an approximation to the xc bias functional.
Numerical results obtained with this functional demonstrate that
the doping-driven Mott transition can indeed be captured with i-DFT.
More generally, our i-DFT approach to calculate spectral functions establishes 
an explicit form for the functional relationship between the many-body spectral 
function and the ground state electronic density. 
}}

\vspace{\baselineskip}

\noindent\textcolor{white!90!black}{%
\fbox{\parbox{0.975\linewidth}{%
\textcolor{white!40!black}{\begin{tabular}{lr}%
  \begin{minipage}{0.6\textwidth}%
    {\small Copyright attribution to authors. \newline
    This work is a submission to SciPost Physics. \newline
    License information to appear upon publication. \newline
    Publication information to appear upon publication.}
  \end{minipage} & \begin{minipage}{0.4\textwidth}
    {\small Received Date \newline Accepted Date \newline Published Date}%
  \end{minipage}
\end{tabular}}
}}
}


\vspace{10pt}
\noindent\rule{\textwidth}{1pt}
\tableofcontents
\noindent\rule{\textwidth}{1pt}
\vspace{10pt}


\section{Introduction}
\label{sec:intro}

A Mott insulator is a material that according to conventional band theory would be classified as 
a metal~\cite{BoerVerwey:37}, but due to strong electronic interactions a gap ---the Mott gap--- 
opens in the electronic spectrum, thus turning it into an insulator~\cite{MottPeierls:37}.
The insulating behavior is therefore driven by the interactions between electrons rather
than band-filling, as in the case of a normal band insulator.
Hence, by tuning the interaction strength a phase transition between a metallic and an insulating 
phase, known as the Mott (metal-insulator) transition, can be induced~\cite{Mott:68}.
On the other hand, a Mott insulator can also be turned into a metal by doping, e.g., by tuning the 
chemical potential via a gate or by charge transfer from donor atoms, much like doping of a 
semiconductor.
However, different from semiconductors, which become conducting by fractional occupation of the 
conduction band, a doped Mott insulator becomes conducting due to the appearance of a 
quasiparticle band within the Mott gap~\cite{FisherKotliarMoeller:95,WernerMillis:07}, which owing 
to the strong electronic interactions is often strongly 
renormalized~\cite{ZitkoHansenPerepelitskyMravljeGeorgesShastry:13,LoganGalpin:16,Hewson:16}.
The strong renormalization may give rise to so-called heavy-fermion behavior, where the effective 
mass of the electrons can be up to a thousand times the free electron mass, leading to peculiar 
thermodynamic properties of these materials~\cite{Coleman:07}. 
Ultimately, doped Mott insulators are also thought to be at the heart of high-Tc 
superconductivity~\cite{OrensteinMillis:00}.

One of the central models for understanding Mott insulators and strongly correlated materials
in general is the Hubbard model~\cite{Hubbard:63,Gutzwiller:63,Kanamori:63}. 
In its simplest version, i.e., the single-band Hubbard model, it consists of a lattice with a 
single orbital at each lattice site, connected by a hopping $t$ and a strictly local (on-site) 
Coulomb repulsion $U$. 
Despite its relative simplicity the Hubbard model not only captures the essence of the Mott 
metal-insulator transition~\cite{Hubbard:64}, but gives rise to an extremely rich phase diagram 
that even includes a superconducting phase~\cite{RothChenSenguptaGeorges:25}.
A particularly clear and physically transparent picture of the Mott transition is obtained in the 
limit of infinite dimensions and the resultant theoretical framework of dynamical mean-field theory 
(DMFT) in terms of local electronic correlations~\cite{Metzner:PRL:1989,Georges:PRB:1992,Georges:RMP:1996}.
More recently, the Hubbard model has also been successfully employed for the theoretical 
description of magnetic nanographenes~\cite{FernandezRossier:07,Valli:18,Ortiz:19,Jacob:22}.
Thus, much of research in strongly correlated electrons has traditionally concentrated on the 
Hubbard model and its ramifications.

While it has always been clear that density functional theory (DFT), being an in principle exact 
many-body framework, should be capable of describing strongly correlated systems, the application 
of DFT to strongly correlated materials has been hampered by two serious caveats: 
(i) the common approximations to the exchange-correlation (xc) functional like LDA and GGA do not 
properly capture the strong electronic correlations that open the Mott gap, and (ii) the Kohn-Sham 
(KS) spectra of a DFT calculation do not provide a reliable description of the actual many-body 
spectra.
With regard to the former, it turns out that an essential ingredient for capturing the strong 
electronic correlations leading to the opening of the Mott gap within 
DFT~\cite{LimaSilvaOliveiraCapelle:03} is the notorious derivative discontinuity of exact 
DFT~\cite{PerdewParrLevyBalduz:82}, which is lacking in the standard approximations like 
LDA and GGA.
On the other hand, even if we had the exact functional or a proper approximation featuring a 
derivative discontinuity, the KS system as an effective \emph{one-body} system cannot properly 
reproduce the many-body spectrum of a Mott insulator, since it must be metallic by construction.
In actual fact, the exact band gap of a system is really the sum of the KS band gap and the 
derivative discontiuity. 
Thus, for a Mott insulator the band gap is entirely due to the derivative discontinuity, which 
offers an alternative definition of a Mott insulator.

To date most efforts to describe strongly correlated materials on an ab initio basis
concentrate on combining DFT calculations with DMFT calculations for a strongly correlated 
subset of orbitals~\cite{Lichtenstein:PRB:1998,Kotliar:RMP:2006,Karolak:JPCM:2011}.
While this DFT+DMFT approach has been quite successful, the notorious double-counting problem 
inherent to this type of approaches remains essentially unsolved~\cite{Karolak:JESRP:2010}.
Combining a Green's function based ab initio method such as the GW approximation with DMFT solves
this problem, but comes at a considerable computational cost~\cite{Biermann:PRL:2003,Biermann:JPCM:2014}.

Recently, substantial progress has been made in the application of DFT to strongly correlated 
problems.
First, it has been shown that DFT can capture certain aspects of the Kondo effect in the Anderson
impurity model, provided that the exchange-correlation (xc) potential has steps at integer values 
of the impurity occupation, related to the derivative 
discontinuity~\cite{StefanucciKurth:11,BergfieldLiuBurkeStafford:12,TroesterSchmitteckertEvers:12}.
Second, we have shown that the actual many-body spectral function can be extracted from a ground 
state DFT calculation, by making use of steady-state DFT (or i-DFT for brief) in a specific limit, 
called the \emph{ideal STM limit}, where the KS equations for the density and the current decouple 
and the density becomes the equilibrium one~\cite{JacobKurth:18}.
This approach has been successfully applied to the Anderson impurity model, where it correctly 
captures the spectral function, both inside and outside the Kondo regime.
Moreover, the approach can also be applied to calculate spectra of bulk systems, for example, to 
the Hubbard model, where it correctly describes the metal-insulator transition in the spectral 
function~\cite{JacobStefanucciKurth:20}.

The central ingredient for computing the spectra in the i-DFT framework is the xc bias, i.e., the 
xc contribution to the applied bias voltage, which is a functional of the density and the current.
In order to correctly capture both the Kondo peak in the Anderson impurity model, as well as the 
quasiparticle peak in the metallic phase of the Hubbard model, the electron density and spectral 
function need to comply with Fermi-liquid theory~\cite{Nozieres:JLTP:1974,Nozieres:book:1998,Hewson:book:1997}.
In our previous work we exploited this connection to derive exact conditions on the xc bias from 
Fermi-liquid theory for describing the Mott transition in the Hubbard model at particle-hole 
symmetry~\cite{JacobStefanucciKurth:20}. 

Here, we are going to generalize and extend the Fermi-liquid conditions on the xc bias to obtain 
local spectral functions in the absence of particle-hole symmetry. 
Importantly, we show a simpler and more transparent way of deriving these
conditions than in our previous work~\cite{JacobStefanucciKurth:20}, thus
establishing a clear connection between Fermi-liquid theory and i-DFT.
While these conditions are rather general and could be useful for other
models as well, we here apply them
to construct an exchange-correlation bias for the 
Hubbard model away from half-filling, which in turn is applied to calculate spectral functions in
the doped Mott phase. Our results show that this approach correctly describes the doping-induced 
Mott transition in the spectral function of the single-band Hubbard model. 

\section{Model and methodology}

\subsection{The Hubbard model for infinite coordination}

We consider the single-band Hubbard model on some lattice given by the Hamiltonian
\begin{equation}
    \mathcal{H} = t \sum_{\langle i,j  \rangle,\sigma} \left(c_{i\sigma}^\dagger c_{j\sigma} + c_{j\sigma}^\dagger c_{i\sigma} \right)
    + v \sum_i n_i
    + U \sum_i n_{i\up} n_{i\dn}
    \label{hamil}
\end{equation}
where $t$ is the hopping between neighboring sites, $v$ is the gate potential (or on-site energy), and
$U$ is the on-site Coulomb repulsion. Indices $i,j$ refer to lattice sites and $\sigma=\up,\dn$ to spin.
$\langle i,j \rangle$ denotes a sum over pairs $i,j$ of lattice sites restricted to nearest-neighbors.
$c_{i\sigma}$ ($c_{i\sigma}^\dagger$) destroy (create) an electron at site $i$ with spin $\sigma$,
$n_{i\sigma}=c_{i\sigma}^\dagger c_{i\sigma}$ is the number operator of site $i$ and spin $\sigma$,
and $n_i=n_{i\up}+n_{i\dn}$.

The solution for the many-body problem given by the Hubbard Hamiltonian (\ref{hamil})
can be written in terms of the one-electron Green's function (GF) as
\begin{equation}
    G_{\bm{k}}(\omega) = \frac{1}{\omega-v-\varepsilon_{\bm{k}}-\Sigma_{\bm{k}}(\omega)}
    \label{Gk}
\end{equation}
where $\varepsilon_{\bm{k}}$ is the dispersion relation of the lattice and $\Sigma_{\bm{k}}(\omega)$
is the proper self-energy.

A considerable simplification for the solution of the Hubbard model is achieved in the limit of infinite
coordination number of the lattice, $K\to\infty$~\cite{Metzner:PRL:1989,Georges:PRB:1992}.
In this limit electronic correlations become
strictly local, i.e., the self-energy becomes $\bm{k}$-\emph{independent}, $\Sigma_{\bm{k}}(\omega)\to\Sigma(\omega)$.
By providing a reference solution for the Hubbard model, this purely theoretical limit, plays a role similar to the
role played by the electron gas in the context of DFT. Moreover, in real materials the atoms often have a relatively
high coordination ($K\sim12$ for the fcc lattice), so that the assumption of negligible non-local
correlations often provides a reasonable approximation.
In this limit we can write the \emph{local} GF as
\begin{equation}
    G(\omega) = \sum_{\bm{k}} G_{\bm{k}}(\omega) = \sum_{\bm{k}}\frac{1}{\omega-v-\varepsilon_{\bm{k}}-\Sigma(\omega)}
    \label{Gloc}
\end{equation}
Correspondingly, the local spectral function is given by
\begin{equation}
    A(\omega) = \mathrm{i}(G^\ast(\omega)-G(\omega)) = -2\,\Im\, G(\omega)
    = -2 \sum_{\bm{k}} \Im\frac{1}{\omega-v-\varepsilon_{\bm{k}}-\Sigma(\omega)}
\end{equation}
Note that in this definition the spectral function is normalized as $\int\,d\omega\,A(\omega)=2\pi$.

Neglecting non-local correlations, i.e., $\Sigma_{\bm{k}}(\omega)\approx\Sigma(\omega)$,
allows to map the strongly correlated lattice problem onto a self-consistent
local Anderson impurity model (AIM) problem.
This is the essence of dynamical mean-field theory (DMFT)~\cite{Georges:RMP:1996,Vollhardt:JPSJ:2005,Kotliar:RMP:2006}. 
The AIM can then be solved very accurately for example by the numerical
renormalization group (NRG)~\cite{Bulla:RMP:2008} or continuous-time quantum Monte-Carlo (CTQMC)~\cite{Gull:RMP:2011} methods.
While we will not use DMFT here to solve the Hubbard model (\ref{hamil}), below we
make use of the specific structure of the local GF in DMFT (\ref{Gloc}) together
with Fermi liquid conditions to derive exact conditions for constructing exchange
correlation potentials for i-DFT.

For the lattice we consider the Bethe lattice in the limit of infinite coordination,
$K\to\infty$. In order for the bandwidth $W$ to remain finite in this limit, the hopping
$t$ needs to scale as $t=t^\ast/\sqrt{K}$, where $t^\ast$ is a constant. The non-interacting
spectral function for the Bethe lattice for gate $v$ is then described 
by a semi-circle of width $W=4t^\ast$:
\begin{equation}
    A_0(\omega-v) =  \frac{4}{D^2} \sqrt{ D^2 - (\omega-v)^2 }
    \label{sf}
\end{equation}
where $D=W/2=2t^\ast$ is the half-bandwidth.
In the following we take $t^\ast$ as the unit of energy.

\subsection{Treating the model with ground state DFT: construction of an xc potential}



In a first step before aiming for the spectral function, we want to treat the
system of interacting electrons on the Bethe lattice with infinite
coordination described by the Hamiltonion (\ref{hamil}) within DFT. Since
we are interested in the uniform system, i.e., the lattice potential $v$ is the
same for all sites of the lattice, by symmetry the density $n$ on a lattice site
also must be the same on all sites. The main idea of DFT is to reproduce the
interacting density as the density of a fictitious, non-interacting system
of electrons, the Kohn-Sham (KS) system, subject to an effective single-particle
KS potential. For the uniform case, by symmetry the KS potential $v_s$ is
the same for all lattice sites and may be written as
\begin{equation}
    v_s = v + v_{\rm Hxc}(n)
    \label{kspot}
\end{equation}
where $v_{\rm Hxc}$ is the Hartree-exchange-correlation (Hxc) potential which
depends on the density. Since the KS system is a system of non-interacting electrons, the
corresponding KS spectral function (or local density of states (LDOS)) is given by  
$A_s(\omega)=A_0(\omega-v_s)$ with $A_0$ defined in Eq.~(\ref{sf}). As usual, the
density can be obtained from the LDOS via
\begin{equation}
    n = 2 \int \frac{{\rm d}\omega}{2 \pi} \;f_T(\omega)\, A_s(\omega)
    \label{dens}
\end{equation}
with the Fermi function $f_T(\omega)=(1+\exp(\beta \omega))^{-1}$ and the
inverse temperature $\beta=1/T$. In the limit of zero temperature, the
frequency integral in (\ref{dens}) can be evaluated to give
\begin{equation}
    n = 1 - \frac{2}{\pi}\arcsin\left(\frac{v_s}{D}\right)
    - \frac{2}{\pi}\cdot \frac{v_s}{D} \sqrt{1 - \left(\frac{v_s}{D}\right)^2}
    \label{dens2}
\end{equation}
Since the KS potential $v_s$ depends on the density via the Hxc potential
(see Eq.~(\ref{kspot})), Eq.~(\ref{dens2}) is the KS self-consistency condition
for the density which has to be solved to obtain the density for a given
external potential $v$. In practice, this means that we need to employ an
approximation for $v_{\rm Hxc}(n)$. In the following, we will construct such
an approximate Hxc potential using known results for our model for the
interacting case.

In Ref.~\cite{LoganGalpin:16}, Logan and Galpin investigate the density as function
of both the interaction strength and the on-site potential using DMFT+NRG. 
Their most important result
is that for interactions $U$ larger than a critical value $U_c$ there exists a whole
range of on-site potentials for which the density is essentially constant at half-filling. 
Physically, this region corresponds to the Mott insulating phase. In the context of
DFT it is known \cite{StefanucciKurth:11,KurthStefanucci:17} that these plateaus in the
density are related to step features in the Hxc potential at half filling ($n=1$) where the
height of the step is basically given by the notorious xc contribution $\Delta_{\rm xc}$ to
the derivative discontinuity of DFT \cite{PerdewParrLevyBalduz:82}. In our attempt to
parametrize the Hxc potential of our model, we will therefore first concentrate on the
work of Ref.~\cite{LoganGalpin:16} to extract $\Delta_{\rm xc}$. We define the parameter 
\begin{equation}
\eta(U) = 1 + \frac{2 v}{U} \;.
\end{equation}
For any value of $U$, $\eta=0$ corresponds to the particle-hole symmetric point 
with density $n=1$. For $\eta>0$ ($\eta<0$) one has $n\leq 1$ ($n\geq 1$). For 
interactions $U>U_c$ and for on-site potentials above the particle-hole symmetric point
($v_{\rm ph}=-U/2$), Logan and Galpin define $\eta_c(U)$ as the line separating the Mott insulating
phase from the metallic phase. If we write the on-site potential corresponding to $\eta_c >0$ as
$\bar{v}^> = - U/2 + v_c$ (with $v_c>0$), then by particle-hole symmetry, the line separating
the Mott and metallic phases corresponds to the gate $\bar{v}^<=-U/2-v_c$ and $-\eta_C(U)$. 
In terms of the on-site potential, the width of the plateau with density $n=1$ then is given by
$2 v_c$ which corresponds exactly to the xc part $\Delta_{\rm xc}$ of the derivative discontinuity of
DFT, i.e.,
\begin{equation}
    \Delta_{\rm xc} = 2 v_c = U \eta_c(U) 
    \label{delta_xc_eta}
\end{equation}
which is valid for $U > U_c$ and $\Delta_{\rm xc}=0$ for $U\leq U_c$. 
For $U > U_c$, Logan and Galpin obtain an excellent fit for $\eta_c$ to their NRG results by solving
the equation
\begin{equation}
    \tilde{U}-\tilde{U}_c = \exp(\phi) \left( \frac{1}{1-\eta_c(U)} \exp(-\phi (1-\eta_c(U))\tilde{U})
    - \exp(-\phi \tilde{U}_c)\right)
    \label{eq_eta_c}
\end{equation}
where $\tilde{U}=U/D$ and $\phi=8$ is a parameter. For $U_c$, the value $U_c=5.88$ has been calculated
using NRG in Ref.~\cite{Bulla:99}.

Having established a relation for the xc part $\Delta_{\rm xc}$ of the derivative discontinuity in terms
of known many-body results, a reasonable ansatz for the Hxc potential can be given as
\begin{equation}
v_{\rm Hxc}(n) = v_{\rm stp}(\Delta_{\rm xc},n-1) + (U-\Delta_{\rm xc}) \left(g(n-1) + \frac{1}{2} \right)
\label{vhxc}
\end{equation}
Here $v_{\rm stp}(\Delta_{\rm xc},x)$ is a step function of step height $\Delta_{\rm xc}$ which for $x<0$
essentially vanishes while for $x>0$ it essentially takes on the value $\Delta_{\rm xc}$. Here we
deliberately say ``essentially'' because for the KS self-consistency condition to have a solution for
all possible on-site potentials $v$, we need the Hxc potential to be continuous everywhere.
This step-like but mathematically continuous feature
  of the Hxc potential then leads to pinning of the density to half-filling for
  a range of on-site potentials~\cite{StefanucciKurth:11}.
In practice, we either take $v_{\rm stp}(\Delta_{\rm xc},x) = \Delta_{\rm xc} f(-x)$ with the Fermi function
$f(-x)$ evaluated at very low temperature or, alternatively, we use the low-temperature limit of the
single-site model Hxc potential given in Eq.~(6) of Ref.~\cite{StefanucciKurth:11}. 

For the function $g(x)$ we impose the symmetry condition $g(-x)=-g(x)$ due to particle-hole symmetry
and require that $g(1)=1/2$. This latter condition together with particle-hole symmetry ensures the
physical constraints that at $n=0$ the Hxc potential strictly vanishes while at full occupation $n=2$
the Hxc potential takes the value $U$. Of course, the strict purpose of a parametrization of $v_{\rm Hxc}$
(and thus of the function $g$) is to reproduce the density of the interacting system by means of the
effectively non-interacting KS system. However, $v_{\rm Hxc}$ will also play an important role
in the construction of the xc bias of i-DFT needed to compute spectral functions (see
Sec.~\ref{xcbias_param}). Therefore when searching for an appropriate parametrization of the function
$g(x)$ we were inspired by earlier work \cite{JacobStefanucciKurth:20} where we parametrized the xc bias
for our model (restricted to the particle-hole symmetric point) which was able to properly describe
the Mott transition. We finally settled on the following form
\begin{equation}
    g(x) = (1-b) \sgn(x) a_2(U) \left( \sqrt{a_1(U) + \tfrac{1}{4}|x|} - \sqrt{a_1(U)} \right) +
\frac{b}{2} x
\label{gfunc}
\end{equation}
where $b=-1/4$ and $\sgn(x)$ is the sign function. The parameter $a_1(U)$ was introduced to both avoid
the singularity of the derivative of the square root and chosen such as to best reproduce the
NRG densities as function of gate given in Fig.~3 of Ref.~\cite{Hewson:16}. It is parametrized as
\begin{equation}
a_1(U) =  0.0697 \exp(-1.45 U) \;.
\label{a1}
\end{equation}
On the other hand, $a_2(U)$ is given as
\begin{equation}
a_2(U) = \frac{1}{2 (\sqrt{1/4 + a_1(U)}-\sqrt{a_1(U)})} 
\label{a2}
\end{equation}
and was chosen in this way to ensure that $g(1)=1/2$ and $g(-1)=-1/2$.

\begin{figure}
  \includegraphics[width=\linewidth]{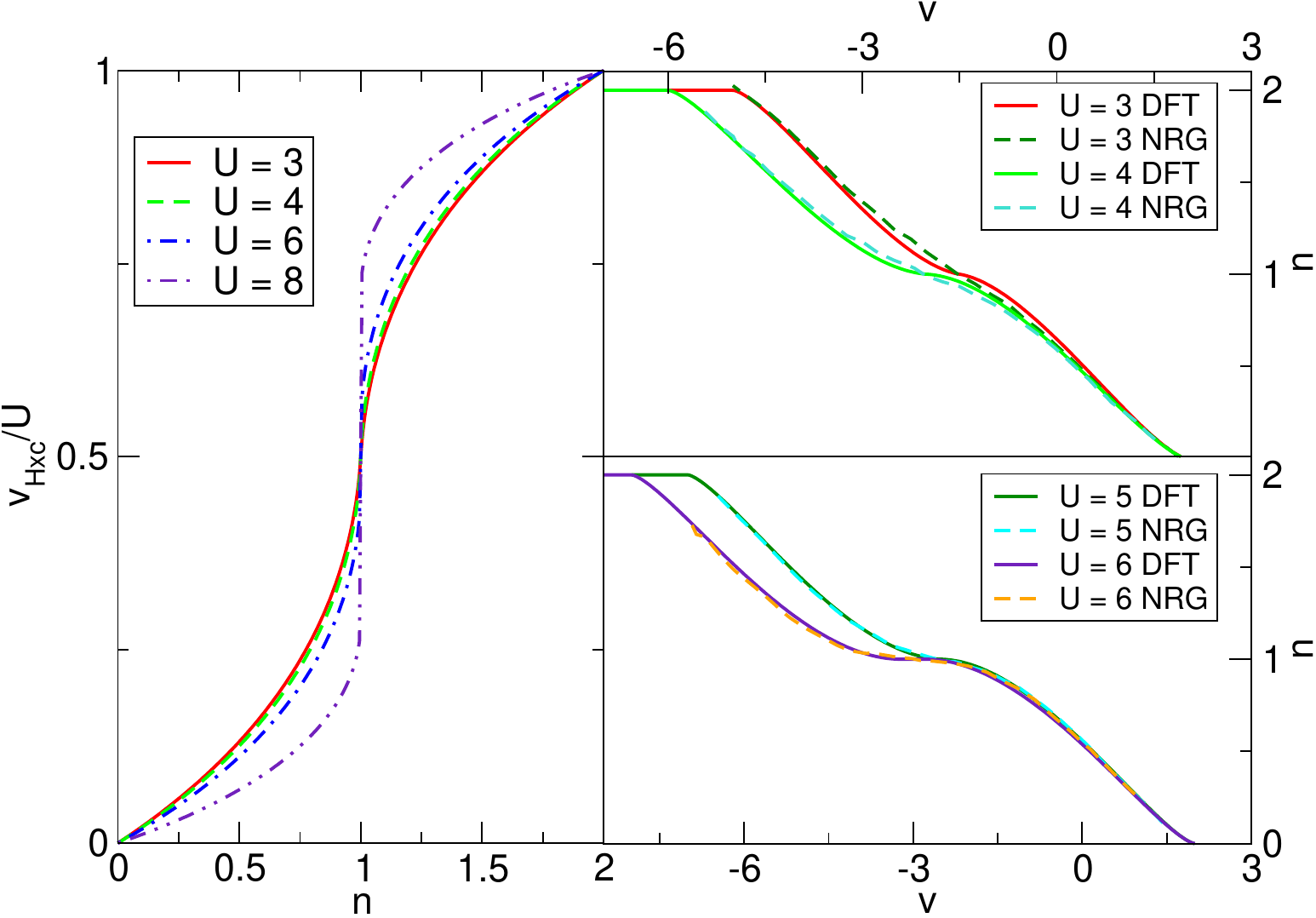}
  \caption{Left panel: Hxc potentials $v_{\rm Hxc}(n)$ of Eqs.~(\ref{vhxc}) and (\ref{gfunc}) in units
    of $U$ for different values of $U$. Right panels: comparison of DFT densities obtained with our
    parametrization of $v_{\rm Hxc}$ with NRG densities of Ref.~\cite{Hewson:16}. Upper right panel:
    $U=3$ and $U=4$, lower right panel: $U=5$ and $U=6$.}
  \label{fig:vhxc_dens}
\end{figure}

In the left panel of Fig.~\ref{fig:vhxc_dens} we show our parametrized Hxc potentials for a few values
of the interaction $U$ (note the sharp steps for $U>U_c$). In the right panels of the same figure we
compare densities as function of the on-site potential $v$ obtained with DFT for various values of $U$
with reference NRG results from Ref.~\cite{Hewson:16}. We note that for smaller values of $U$ the
DFT results show small differences to the NRG ones. However, for larger values of the interaction, DFT and
NRG densities are hardly distinguishable on the scale of the plot.

\section{Density functional approach to doping-driven Mott transition}

\subsection{Synopsis of steady-state density functional theory}
\label{sec:idft}
  
Steady-state density functional theory, also called i-DFT 
\cite{StefanucciKurth:15}, is a generalization of standard ground-state (or 
equilibrium) DFT to describe {\em steady-state} electronic transport through 
an arbitrary spatial region connected to two biased leads. It is based on a
one-to-one correspondence between the local potential in the arbitrary region
plus the bias across it and the corresponding (non-equilibrium) density in the
region plus the steady-state current across it. The corresponding KS system
then gives both the density and the steady current $I$, where not only the KS
potential carries an Hxc contribution, but also the KS bias $V_s$ is
the sum of the external bias $V$ plus an xc contribution to the bias
$V_{\rm xc}$, i.e., $V_s = V + V_{\rm xc}$. Both the Hxc contribution to the
external potential as well as the xc bias are functionals of both density and
current. In general, the KS equations for these two basic variables are 
coupled and have to be solved together self-consistently. 

In the following, we will use a very particular limit of i-DFT (or any theory
of electronic transport) which in Ref.~\cite{JacobKurth:18} was called
the ``ideal STM limit'' because it resembles the way a scanning tunneling
microscope (STM) works. Assume that the following conditions are met: (i) one of the
leads (which is kept at zero temperature) is coupled infinitesimally weakly to
the interacting region of interest and (ii) the whole applied bias drops at this
lead. Then the differential conductance for a given bias $V=\omega$ is related
to the zero-temperature {\em equilibrium} spectral function of the region
connected to the second lead, i.e.,
\begin{equation}
\frac{\partial I}{\partial V}\bigg\vert_{V=\omega}
\xrightarrow[{\mathbf \Gamma_T} \to 0]{} \frac{1}{\pi}
\Tr[{\mathbf{\Gamma_T A}}(\omega)]
\label{specfunc_mb}
\end{equation}
where ${\mathbf \Gamma_T} = i ( {\mathbf \Sigma_T}^{\dagger} -
{\mathbf \Sigma_T})$ with ${\mathbf \Sigma_T}$ being the embedding self
energy of the tip and ${\mathbf A}(\omega)$ is the equilibrium many-body
spectral function matrix.

In i-DFT, the ideal STM limit leads to a complete decoupling of the
self-consistent equation for the density and the one for the current
\cite{JacobKurth:18}. This means that one can first solve the KS for the
equilibrium density and then, for this density, solve the equation for the
current. This latter equation reads
\begin{equation}
    I = \frac{1}{\pi} \int {\rm d} \omega\; (f_{T=0}(\omega-V_s) - f_T(\omega))
    \; \Tr[\mathbf{\Gamma}_T \mathbf{A}_s(\omega)]
    \label{current} 
\end{equation}
where ${\mathbf A}_s(\omega)$ is the KS spectral function matrix. The coupling
to the ``STM tip'', i.e., the weakly coupled electrode, can be chosen at will.
As we are interested in the local spectral function of our model, we choose
the coupling matrix $\mathbf{\Gamma}_T$ such that the tip only couples to the 
single site $j$ (for our model all sites are equivalent) such that
$(\mathbf{\Gamma}_T)_{jk} = \gamma_T \delta_{jk}$ and we take $\gamma_T$ to be
independent of frequency, i.e., we use the wide-band limit for the tip. As a
final ingredient we realize that the above procedure can also be applied to
an infinitely extended bulk system connected only to the STM
tip \cite{JacobStefanucciKurth:20}. In this case, the bulk takes on the role
of the second electrode. 

Putting everything together and performing the derivative of
Eq.~(\ref{current}) with respect to the bias, in Ref.~\cite{JacobKurth:18} we
arrived at the following compact relation which expresses the (local)
many-body spectral function completely in terms of KS quantities as
\begin{equation}
    A(\omega) = \frac{A_s(\omega + V_{\rm xc})}{1 -
    \left.\frac{\partial V_{\rm xc}}{\partial \tilde{I}}\right|_{n,\It(\omega)} A_s(\omega + V_{\rm xc})} 
    \label{specfunc_idft}
\end{equation}
where we have defined the reduced current as $\tilde{I}= \pi I/\gamma_T$.
In Eq.~(\ref{specfunc_idft}) both the xc bias and its derivative have to be
evaluated at the self-consistent current corresponding to the external bias
$V=\omega$.
Realizing that the KS spectral function depends on the density $n(\bm{r})$
via the KS potential $v_s(\bm{r})$, we see that Eq.~(\ref{specfunc_idft}) actually
presents a \emph{functional} relationship between the ground state density $n(\bm{r})$ 
and the spectral function via the xc bias. Note that the current $I=I(\omega)$ is merely an 
auxiliary quantity for determining the spectral function. Thus Eq.~(\ref{specfunc_idft})
expresses the fact that in DFT the exact spectral function is a functional of the 
ground state density, and defines that functional in terms of the xc bias.

Of course, in order to apply the scheme described here to the calculation of
many-body spectral functions one needs an approximation for the xc bias
$V_{\rm xc}$. In the following, we first derive relations for this quantity
from Fermi liquid theory. Using these relations we then construct an
approximation for $V_{\rm xc}$ for the model under study. 
Introducing, as in Ref.~\cite{JacobStefanucciKurth:20} the function $\Omega(\omega)\equiv\omega+V_\xc(\It(\omega))$ we can rewrite 
Eq.~(\ref{specfunc_idft}) as
\begin{equation}
    \label{specfunc_idft_omega}
    A(\omega) = A_s(\Omega)\,\frac{d\Omega}{d\omega}
\end{equation}
which will be exploited below a couple of times in the derivation of the Fermi liquid conditions for the xc bias.

\subsection{Fermi liquid conditions for the exchange-correlation bias}

We now make use of Fermi liquid theory~\cite{Nozieres:book:1998,Hewson:book:1997} and the local GF (\ref{Gloc}) in order to 
derive some exact conditions for the exchange-correlation bias.
First, we derive relationships between the exact spectral function $A(\omega)$
and the KS spectral function $A_s(\omega)$ as well as their derivatives.

Expanding the local self-energy to first order (purely real) 
\begin{equation}
    \Sigma(\omega) \approx \Sigma(0) + \Re\,\Sigma^\prime(0)\,\omega
\end{equation}
and defining the quasiparticle weight
\begin{equation}
    Z = (1-\Re\,\Sigma^\prime(0))^{-1}
    \label{qpweight}
\end{equation}
we can approximate the local (many-body) GF close to the Fermi level as:
\begin{equation}
    G(\omega) \approx \sum_{\bm{k}} \frac{Z}{\omega-Z\,(v+\Re\,\Sigma(0)) - Z\,\varepsilon_{\bm{k}}}
    = \sum_{\bm{k}} \frac{Z}{\omega-Z\,v^\ast - Z\,\varepsilon_{\bm{k}}} \; ,
    \label{G}
\end{equation}
where in the last step we have introduced the effective gate $v^\ast$ given by
the actual gate $v$ and the Hartree-shift $\Re\,\Sigma(0)$, $v^\ast=v+\Re\,\Sigma(0))$. 

On the other hand, the local Kohn-Sham GF is given by:
\begin{equation}
    G_s(\omega) = \sum_{\bm{k}} \frac{1}{\omega-v_s -\varepsilon_{\bm{k}}} \; .
    \label{Gs}
\end{equation}

The first condition we derive follows from the Friedel sum rule and relates the 
many-body GF (\ref{G}) to the Kohn-Sham one (\ref{Gs}) evaluated at the Fermi level ($\omega=0$):
\begin{equation}
    \label{FSR}
    \sum_{\bm{k}} \frac{1}{v^\ast + \varepsilon_{\bm{k}}}
    = \sum_{\bm{k}} \frac{1}{v_s + \varepsilon_{\bm{k}}} \;,
\end{equation}
where the quasiparticle weight $Z$ cancels out on the l.h.s.
This implies that the effective gate $v^\ast$ must be equal to the Kohn-Sham gate $v_s$,
\begin{equation}
    v^\ast = v+\Sigma(0) = v_s \; ,
\end{equation}
We can thus write the many-body GF as:
\begin{equation}
    G(\omega) = \sum_{\bm{k}} \frac{Z}{\omega-Z\,v_s - Z\,\varepsilon_{\bm{k}}} \;.
    \label{G2}
\end{equation}
Taking the first and second derivative of the many-body GF (\ref{G2}), we obtain:
\begin{eqnarray}
    \label{Gp}
    G^\p(\omega) &=& -\sum_{\bm{k}} \frac{Z}{(\omega-Z\,v_s - Z\,\varepsilon_{\bm{k}})^2} \\
    \label{Gpp}
    G^\pp(\omega) &=& 2\sum_{\bm{k}} \frac{Z}{(\omega-Z\,v_s - Z\,\varepsilon_{\bm{k}})^3}
\end{eqnarray}
On the other hand, the derivatives of the KS GF yield:
\begin{eqnarray}
    \label{Gsp}
    G_s^\p(\omega) &=& -\sum_{\bm{k}} \frac{1}{(\omega - v_s - \varepsilon_{\bm{k}})^2} \\
    \label{Gspp}
    G_s^\pp(\omega) &=& 2\sum_{\bm{k}} \frac{1}{(\omega - v_s - \varepsilon_{\bm{k}})^3}
\end{eqnarray}
Evaluating the derivatives at the Fermi level the following aditional relationships between
the many-body and KS GFs arise:
\begin{eqnarray}
    G^\p(0)  &=& -\sum_k \frac{Z^{-1}}{(\epsilon_k+v_s)^2} = Z^{ -1} \, G_s^\p(0) \\
    G^\pp(0) &=& -\sum_k \frac{2\,Z^{-2}}{(\epsilon_k+v_s)^3} = Z^{ -2} \, G_s^\pp(0)
\end{eqnarray}
These relationships are a direct consequence of Friedel sum rule.
In terms of the spectral functions these relationships can be summarized as:
\begin{empheq}[box=\fbox]{align}
    \label{FL-A0}
    \hspace{1ex} A(0)     &= A_s(0) \hspace{1ex}  \\
    \label{FL-Ap0}
    \hspace{1ex} A^\p(0)  &= Z^{-1} \, A^\p_s(0) \hspace{1ex} \\
    \label{FL-App0}
    \hspace{1ex} A^\pp(0) &= Z^{-2} \, A^\pp_s(0) \hspace{1ex} 
\end{empheq}

Using the Fermi-liquid conditions (\ref{FL-A0}-\ref{FL-App0}) between the many-body and KS spectral functions and their derivatives,
we now derive conditions for the xc bias. 
Additionally, we will exploit the exact i-DFT relationship between the KS 
and many-body spectral functions (\ref{specfunc_idft_omega}).

The first condition simply follows from charge conservation, i.e., 
the KS system should reproduce the total charge of the real system, thus
\begin{equation}
n = \int_{-\infty}^0\,d\omega\;A(\omega) = \int_{-\infty}^0\,d\omega\;A_s(\Omega(\omega))\,\frac{d\Omega}{d\omega} 
= \int_{-\infty}^{\Omega(0)}\,d\Omega\;A_s(\Omega) \stackrel{(!)}{=} \int_{-\infty}^{0}\,d\omega\;A_s(\omega) \;.
\end{equation}
It follows that $\Omega(0)$ must vanish, and hence so must the xc bias $V_\xc$ for zero current:
\begin{equation}
    \label{cond1}
    \Omega(\omega=0) = 0+V_\xc(n,\It(0)) \; \Rightarrow \; V_\xc(n,0)=0 \;.
\end{equation}
The next condition follows directly from the Friedel sum rule:
\begin{equation}
    A_s(\omega=0) \stackrel{(!)}{=} A(\omega=0) = A_s(\Omega(0)) \, \left.\frac{d\Omega}{d\omega}\right|_0 
    = A_s(0) \, \left( 1 + \left.\frac{\partial V_\xc}{\partial\It}\right|_{n,0} \, A(0) \right) \;,
\end{equation}
where in the last step we have used the chain rule and $A(\omega)=d\It/d\omega$.
Therefore, the quantity in parenthesis must equal 1, and we arrive at the second FL condition for the xc bias, 
which states that its first derivative w.r.t. the current must vanish at zero current:
\begin{equation}
    \label{FL2-xc}
    \left.\frac{\partial V_\xc}{\partial \It}\right|_{n,0} = 0 \;. 
\end{equation}
Next we use the FL condition (\ref{FL-Ap0}) to derive a condition for the second derivative of the xc bias:
\begin{equation}
    Z^{-1} A_s^\p(0) = A^\p(0) = \left.\frac{d}{d\omega}\right|_0 \left( A_s(\Omega(\omega))\frac{d\Omega}{d\omega} \right) 
    = \left[ A_s^\p\left(\Omega(\omega)\right) \left( \frac{d\Omega}{d\omega} \right)^2 + A_s\left(\Omega(\omega)\right)\,
    \frac{d^2\Omega}{d\omega^2} \right]_{\omega=0}
    \label{interm}
\end{equation}
Using $\Omega(0)=0$ and $\Omega^\p(0)=1$, derived above, and solving for $\Omega^\pp(0)$, we obtain:
\begin{equation}
    \label{d2Om0}
    \left.\frac{d^2\Omega}{d\omega^2}\right|_0 = \frac{A_s^\p(0)}{A_s(0)} \left( Z^{-1} - 1 \right)
\end{equation}
On the other hand, using $d\Omega/d\omega=1 + A(\omega)\,\partial V_\xc/\partial\It$, $A(\omega)=d\It/d\omega$, and again the chain rule, we obtain:
\begin{equation}
    \label{d2Om-d2Vxc}
    \frac{d^2\Omega}{d\omega^2} = A^\p(\omega) \frac{\partial V_\xc}{\partial\It} + [A(\omega)]^2 \frac{\partial^2V_\xc}{\partial\It^2}
\end{equation}
Evaluating this expression at $\omega=0$, using the above derived conditions (\ref{FL-A0}) and (\ref{FL2-xc}),
and solving for $(\partial^2V_\xc/\partial\It^2)_{n,0}$, yields the third FL condition for the xc bias:
\begin{equation}
    \label{cond3}
    \left.\frac{\partial^2V_\xc}{\partial\It^2}\right|_{n,0} = \frac{A_s^\p(0)}{A_s(0)^3} \, \left( Z^{-1} - 1 \right) 
\end{equation}

Finally, we derive a condition for the third derivative of the xc bias from FLT.
Starting from (\ref{FL-App0}) and the intermediate result (\ref{interm}), we obtain:
\begin{eqnarray}
    Z^{-2} \, A_s^\pp(0) &=& A^\pp(0) = \frac{d}{d\omega}
    \left[ A_s^\p(\Omega)\,\left( \frac{d\Omega}{d\omega} \right)^2 + A_s(\Omega)\,\frac{d^2\Omega}{d\omega^2} \right]_{\omega=0}
    \nonumber\\
    &=& A_s^\pp(0) + 3\,A_s^\p(0)\,\left.\frac{d^2\Omega}{d\omega^2}\right|_0 + A_s(0)\,\left.\frac{d^3\Omega}{d\omega^3}\right|_0
\end{eqnarray}
where in the last step we have also used $\Omega(0)=0$ and $d\Omega/d\omega(0)=1$.
Solving for $d^3\Omega/d\It^3$ and using the result (\ref{d2Om0}) yields:
\begin{equation}
    \label{d3Om0}
   \left.\frac{d^3\Omega}{d\omega^3}\right|_0 = (Z^{-2}-1)\,\frac{A_s^\pp(0)}{A_s(0)} 
   - 3\,\left( Z^{-1} - 1 \right)\,\left( \frac{A_s^\p(0)}{A_s(0)} \right)^2
\end{equation}
On the other hand, derivation of (\ref{d2Om-d2Vxc}) w.r.t. $\omega$, and applying the chain rule yields
\begin{eqnarray}
    \label{d3Om-d3Vxc}
    \left.\frac{d^3\Omega}{d\omega^3}\right|_0 &=& \left[ 
        A^\pp(\omega)\,\left.\frac{\partial V_\xc}{\partial \It}\right|_{n,\It(\omega)}
        + 3\,A^\p(\omega)\,A(\omega)\,\left.\frac{\partial^2V_\xc}{\partial\It^2}\right|_{n,\It(\omega)} 
        + A(\omega)^3\,\left.\frac{\partial^3V_\xc}{\partial\It^3}\right|_{n,\It(\omega)} 
        \right]_{\omega=0}
        \nonumber\\
        &=& 3\,Z^{-1}\,A_s^\p(0)\,A_s(0)\,\left.\frac{\partial^2V_\xc}{\partial\It^2}\right|_{n,0} 
        + A_s(0)^3\,\left.\frac{\partial^3V_\xc}{\partial\It^3}\right|_{n,0} 
\end{eqnarray}
where we have used $(\partial V_\xc/\partial\It)_{n,0}=0$, $A(0)=A_s(0)$ and $A^\p(0)=Z^{-1}A_s^\p(0)$ in the last step.
Equating (\ref{d3Om0}) and (\ref{d3Om-d3Vxc}), solving for $(\partial^3V_\xc/\partial\It^3)_{n,0}$ and using the third
FL condition (\ref{cond3}), yields the fourth FL condition:
\begin{equation}
    \label{cond4}
    \left. \frac{\partial^3{V_\xc}}{\partial\It^3}\right|_{n,0} = \frac{Z^{-2} - 1}{(A_s(0))^4}
    \left[ A_s^\pp(0) - \frac{3\left(A_s^\prime(0)\right)^2}{A_s(0)} \right]
\end{equation}
In the following we summarize the FL conditions on the xc bias derived above:
\begin{empheq}[box=\fbox]{align}
    \label{FL1}
    \hspace{1ex}  V_\xc(n,0) &= 0 &\mbox{FL-1}\hspace{1ex}\\
    \label{FL2}
    \hspace{1ex} \left.\frac{\partial V_\xc}{\partial\It}\right|_{n,0} &= 0 &\mbox{FL-2}\hspace{1ex} \\
    \label{FL3}
    \hspace{1ex} \left.\frac{\partial^2V_\xc}{\partial\It^2}\right|_{n,0} &= \frac{A_s^\p(0)}{A_s(0)^3} \, \left( Z^{-1} - 1 \right) 
    &\mbox{FL-3} \hspace{1ex} \\ 
    \label{FL4}
    \hspace{1ex} \left. \frac{\partial^3{V_\xc}}{\partial\It^3}\right|_{n,0} &= \frac{Z^{-2} - 1}{(A_s(0))^4}
    \left[ A_s^\pp(0) - \frac{3\left(A_s^\prime(0)\right)^2}{A_s(0)} \right] &\mbox{FL-4}\hspace{1ex} 
\end{empheq}

\subsection{Parametrization of the exchange-correlation bias for arbitrary doping}
\label{xcbias_param}

We now exploit the FL conditions (\ref{FL1}-\ref{FL4}) to construct an approximation for the 
xc bias of the Hubbard model (\ref{hamil}) on the Bethe lattice
with infinite coordination. Extensions to other lattices such as, e.g.,
the single cubic lattice, are expected to follow similar construction
principles but obviously will require different parametrizations for, e.g.,
the Hxc potential or the quasiparticle weight.

\begin{figure}[tb]
  \includegraphics[width=\linewidth]{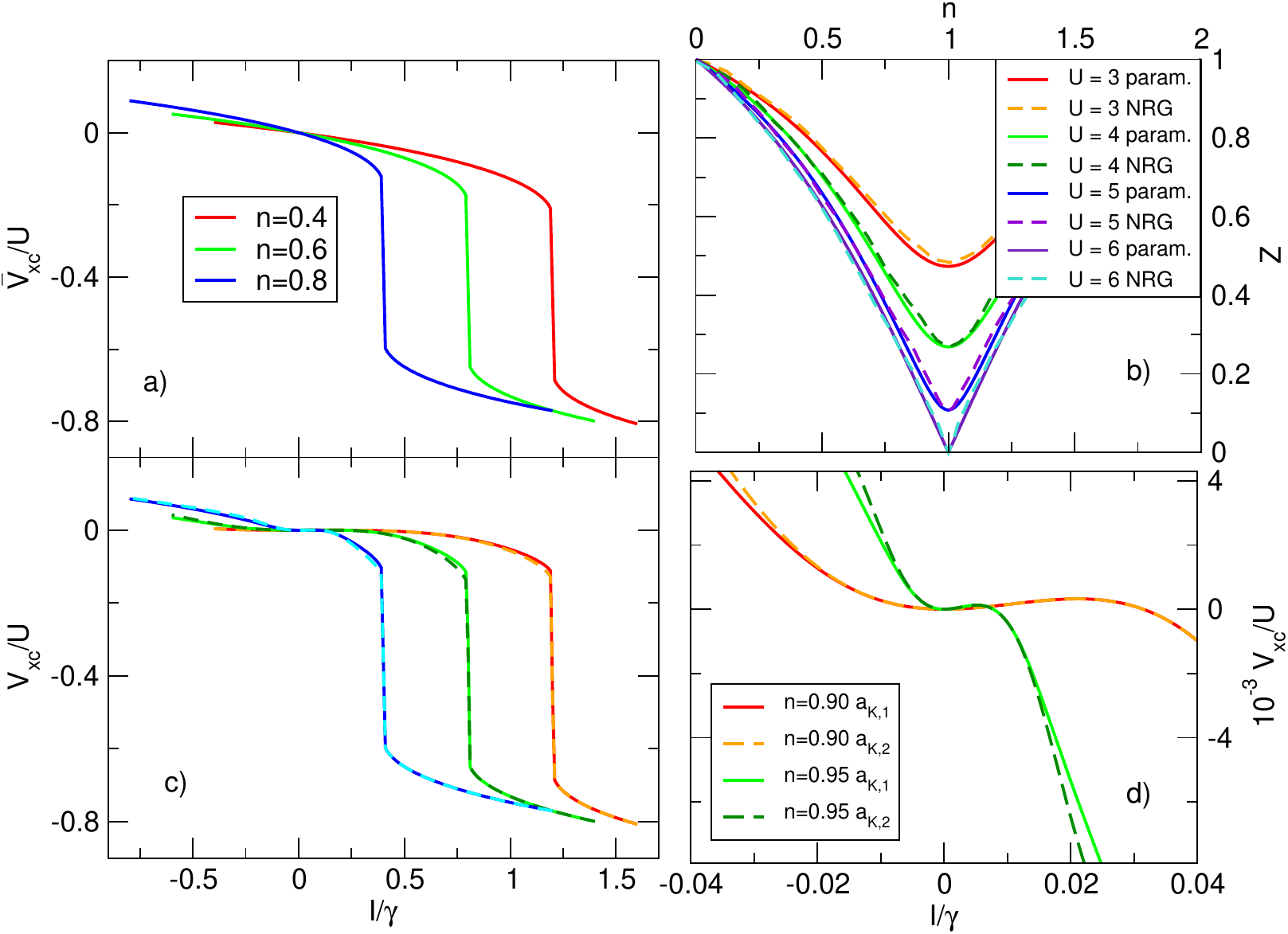}
  \caption{Panels a), c), and d): Exchange-correlation bias $V_{\rm xc}$ for
    $U=8$ in different approximations. Panel a): Mott xc bias according to 
    Eq.~(\ref{mottbias}). Panel c)  xc bias $V_{\rm xc}$ according to 
    Eq.~(\ref{xcbias-ansatz}) with the two different forms of the Kondo 
    prefactor of Eqs.~(\ref{akondo1}) (solid lines) and (\ref{akondo2})
    (dashed lines). Panel d): xc bias with corrected Kondo prefactors. 
    Panel b): quasiparticle weight $Z$ as function of the
    density for various values of $U$ from our parametrization
    (Eqs.~(\ref{z_ph}) - (\ref{b_above_uc})) compared to the NRG results of
    Ref.~\cite{Hewson:16}.}
  \label{fig:qpweight_xcbias}
\end{figure}

Our starting point is the following ansatz for the 
xc bias in the Mott insulating phase in terms of the Hxc potential:
\begin{equation}
    \label{mottbias}
    \bar{V}_\xc(n,\It) = v_\Hxc(n) - v_\Hxc(n+\It/2\pi) \;. 
\end{equation}
This ansatz is motivated by previous experience both with the Anderson impurity model
\cite{StefanucciKurth:15,KurthStefanucci:17,JacobKurth:18},
multiple quantum dots \cite{SobrinoKurth:26}, and with
reverse-engineering of the xc bias in the Mott phase of the Bethe lattice
from dynamical mean field theory in the non-crossing approximation similar
to what has been done in the Supplemental Material of
Ref.~\cite{JacobStefanucciKurth:20}.
By construction (\ref{mottbias}) fulfills FL-1, $\bar{V}_\xc(n,0)=0$.  
Panel a) of Fig.~\ref{fig:qpweight_xcbias} shows the Mott xc bias for different 
values of $n$ for $U=8(>U_c)$. The most prominent feature of 
$\bar{V}_\xc(n,\It)$ is the sharp step of height $\Delta_{\rm xc}$ at 
$I^{\rm stp}/\gamma=2\, (1-n)$. This step is responsible for opening the gap 
in the insulating Mott phase \cite{JacobStefanucciKurth:20}. We also note that 
$\bar{V}_\xc(n,\It)$ is monotonously decreasing with $\It$ which guarantees that 
the corresponding many-body spectral function $A(\w)$ according to 
Eq.~(\ref{specfunc_idft}) is non-negative. Furthermore, due to particle-hole 
symmetry of the 
Hxc potential, $v_{\rm Hxc}(2-n)=U-v_{\rm Hxc}(n)$, we have that 
$\bar{V}_\xc(2-n,-\It)=-\bar{V}_\xc(n,\It)$

Obviously, the second FL condition (\ref{FL2}) is not fulfilled for the Mott 
xc bias (\ref{mottbias}), since $d\bar{V}_\xc/d\It(0)\neq0$ for any $n$.
In order to impose FL conditions (\ref{FL2}-\ref{FL4}) on the xc bias, we follow 
our previous works and introduce a "Kondo prefactor": 
\begin{equation}
    \label{xcbias-ansatz}
    V_\xc(n,\It) = \aK(\It) \, \bar{V}_\xc(n,\It)
\end{equation}
In order to comply with FL-2, the Kondo prefactor must vanish for the zero current, $\aK(0)=0$, since 
\begin{equation}
   0 \stackrel{\rm FL2}{=}
    V_\xc^\prime(n,\It=0)
    = \aK^\prime(0) \, \underbrace{\bar{V}_\xc(n,0)}_{=0} + \aK(0) \, \bar{V}_\xc^\prime(n,0)
    = \aK(0) \, \bar{V}_\xc^\prime(n,0) 
\end{equation}
where the prime denotes derivative with respect to current, i.e., 
$V_\xc^\prime(n,\It)=\partial_{\It}V_\xc(n,\It)$ and $\aK^\prime(\It)=\partial_{\It}\aK(\It)$.
In the last step we have used that $\bar{V}_\xc$ complies with FL1. 

On the other hand, for sufficiently large $|\It|$ the xc bias should recover the form of the Mott xc bias, 
in order to reproduce the lower and upper Hubbard bands, and hence $\aK\to1$. 
This is achieved by choosing the following form for the prefactor:
\begin{equation}
    \label{prefactor}
    \aK(\It) = a( Y(\It) ) \;\mbox{ with }\; Y(\It) = \lambda_1 \It + \lambda_2 \It^2 + \lambda_3 \It^3 + \ldots
\end{equation}
where the envelope function $a(Y)$ is an odd and bounded function, with
$a(Y)\to\pm1$ for $Y\to\pm\infty$. The specific form of $a_{\rm K}$
used in the present work will be given later. 

For the inner function $Y(\It)$ we have chosen a polynomial of finite order $n$, but without a constant term (so that
$Y\to0$ for $\It\to0$). 
We fix the coefficients $\lambda_i$ of the polynomial via the Fermi liquid conditions, thereby enforcing them on the 
xc bias $V_\xc(n,\It)$. Since conditions FL-1 and FL-2 are already enforced by the form of the xc bias given by (\ref{mottbias}) 
and  (\ref{xcbias-ansatz}), we have only two conditions left, FL-3 and FL-4, to fix the coefficients. 
Hence the polynomial must be quadratic, $Y(\It)=\lambda_1\It+\lambda_2\It^2$.
For higher order polynomials we need additional conditions (Fermi liquid or other)
to fix the addional coefficients.

We now exploit conditions FL-3 and FL-4, to fix the coefficients $\lambda_1$ and $\lambda_2$.
Application of condition FL-3 on the xc bias (\ref{xcbias-ansatz}) yields:
\begin{eqnarray}
    V_\xc^\pp(n,\It=0) &=& \aK^\pp(0) \, \underbrace{\bar{V}_\xc(n,0)}_{=0} + 2 \, \aK^\p(0) \, \bar{V}_\xc^\p(n,0) +
    \underbrace{\aK(0)}_{=0} \, \bar{V}_\xc^\pp(n,0) \nonumber\\
    &=&  2 \, \aK^\p(0) \, \bar{V}_\xc^\p(n,0) \stackrel{\rm FL3}{=} (Z^{-1}-1)\,\frac{A_s^\p(0)}{(A_s(0))^3}
\end{eqnarray}
The derivative of the Kondo prefactor yields 
\begin{equation}
    \label{daK0}
    \aK^\p(\It) = a^\p(Y(\It)) \, Y^\p(\It) = a^\p(\It) \, (\lambda_1 + 2\,\lambda_2 \, \It ) \xrightarrow[\It\to0]{} a^\p(0) \, \lambda_1
\end{equation}
Thus, we obtain for the linear coefficient
\begin{equation}
    \lambda_1 = \frac{(Z^{-1}-1)\,A_s^\p(0)}{2\,a^\p(0)\,\bar{V}_\xc^\p(n,0)\,(A_s(0))^3} \;.
\label{lambda1}
\end{equation}
Note that at particle-hole symmetry, we have $A_s^\p(0)=0$, and thus $\lambda_1=0$. Hence
the Kondo prefactor becomes purely quadratic in this case, and thus symmetric in the current,
as it should in order to recover particle-hole symmetry in the i-DFT spectral function.

Finally, we are going to exploit FL4 to fix $\lambda_2$. Application of FL4 to the xc bias (\ref{xcbias-ansatz})
yields:
\begin{eqnarray}
    \label{d3Vxc}
    V_\xc^\ppp(n,\It=0) &=& \aK^\ppp(0)\,\underbrace{\bar{V}_\xc(n,0)}_{=0} + 3\,\aK^\pp(0)\,\bar{V}_\xc^\p(n,0) 
    + 3\, \aK^\p(0)\,\bar{V}_\xc^\pp(n,0) + \underbrace{\aK(0)}_{=0}\,\bar{V}_\xc^\ppp(n,0) \nonumber\\
    &=&  3\,\aK^\pp(0)\,\bar{V}_\xc^\p(n,0) + 3\, \aK^\p(0)\,\bar{V}_\xc^\pp(n,0) 
    \stackrel{\rm FL4}{=} \frac{Z^{-2}-1}{A_s(0)^4} \, \left[ A_s^\pp(0) - \frac{3\,A_s^\p(0)^2}{A_s(0)} \right]  \,.
\end{eqnarray}
The second derivative of the Kondo prefactor yields:
\begin{equation}
    \label{d2aK0}
    \aK^\pp(0) = a^\pp(0) \, \lambda_1 + 2\, a^\p(0) \,\lambda_2 = 2 \, a^\p(0) \,\lambda_2 \,,
\end{equation}
where in the last step we have used that $a(Y)$ is an odd function, and thus $a^\pp(0)=0$.

Plugging this and the result for the first derivative (\ref{daK0}) into (\ref{d3Vxc}), and solving
for $\lambda_2$, we obtain:
\begin{equation}
    \lambda_2 = \frac{Z^{-2}-1}{6\,a^\p(0)\,\bar{V}_\xc^\p(0)\,A_s(0)^4} \left( A_s^\pp(0) - \frac{3\,(A_s^\p(0))^2}{A_s(0)} \right)
    -\frac{\bar{V}_\xc^\pp(0)}{2\,\bar{V}_\xc^\p(0)}\,\lambda_1 \,.
\label{lambda2}
\end{equation}

So far we haven't specified the specific form of the Kondo
prefactor $\aK$ of Eq.~(\ref{xcbias-ansatz}). In previous works
\cite{JacobKurth:18,JacobStefanucciKurth:20} we have chosen the form
\begin{equation}
a_{\rm K,1}(\It)=\tfrac{2}{\pi}\arctan(Y(\It))
\label{akondo1}
\end{equation}
for which also here we will present some results. However, for reasons to be 
explained below we will also use the form 
\begin{equation}
a_{\rm K,2}(\It)=\tanh(Y(\It)) \;.
\label{akondo2}
\end{equation}
With these choices for the Kondo prefactors we have specified the form of our 
approximation to $V_{\rm xc}$. However, another ingredient which needs 
to be specified is the quasiparticle weight $Z$  which enters in the 
definintions of the parameters $\lambda_1$ and $\lambda_2$ of  
Eqs.(\ref{lambda1}) and (\ref{lambda2}), respectively. 
Before going to describe a parametrization for the quasiparticle
weight $Z$ we would also like to highlight how our functional (which uses
input from Fermi liquid theory) transitions to the Mott insulating regime
which cannot be described in this framework: in the Mott regime, $Z$ vanishes
(see below) and therefore the coefficient $\lambda_2$ of Eq.~(\ref{lambda2})
diverges (to $+\infty)$). Then both Kondo prefactors (\ref{akondo1})
and (\ref{akondo2}) become unity and our xc bias functional reduces to
$\bar{V}_{\rm xc}$ which correctly describes the Mott phase.

The quasiparticle weight of our model has been computed as function of the 
density with NRG in Ref.~\cite{Hewson:16} and we now construct a parametrization 
of these results. We start at the particle-hole symmetric point ($n=1$) for which 
the quasiparticle weight can be parametrized as 
\begin{equation}
Z_{\rm ph} = \left( 1 + \frac{\gamma_z^2}{U_c^2}\right)
\left[ \frac{\gamma_z^2}{\gamma_z^2 + U^2} -
  \frac{\gamma_z^2}{\gamma_z^2 + U_c^2} \right]
\label{z_ph}
\end{equation}
with $\gamma_z=4$. Then for $U\leq U_c$ we parametrize the density dependence 
of $Z$ for arbitrary density as 
\begin{equation}
Z = \sqrt{Z_{\rm ph}^2 + \left(1-Z_{\rm ph}^2\right)
  \left(\frac{\arctan(b(n-1))}{\arctan(b)}\right)^2}
\label{z_below_uc}
\end{equation}
with 
\begin{equation}
b = 1.79 - 0.000469 U^4 \;.
\label{b_below_uc}
\end{equation}
For $U>U_c$ we still use the parametrization of Eq.~(\ref{z_below_uc}) but with 
$Z_{\rm ph}=0$ and the slightly different parametrization for $b$ 
\begin{equation}
b = \frac{1.3}{1 + (U-U_c)^2} \;.
\label{b_above_uc}
\end{equation}

In panel b) of Fig.~\ref{fig:qpweight_xcbias} we confirm that our
parametrization of $Z$ reproduces the NRG results of Ref.~\cite{Hewson:16}
reasonably well. With this, in principle we have all the necessary ingredients
to construct an approximate xc bias which includes the Fermi-liquid conditions
derived above. In panel c) we show our approximate $V_{\rm xc}$ according to
the ansatz (\ref{xcbias-ansatz}) for the two forms of the Kondo prefactor
(\ref{akondo1}) and (\ref{akondo2}) with the coefficients $\lambda_1$
and $\lambda_2$ chosen according to Eqs.~(\ref{lambda1}) and
(\ref{lambda2}). Here we choose the densities below $n=1$. For currents
smaller than the one corresponding to the step in $V_{\rm xc}$, our
approximation is somewhat similar to the Mott xc bias
$\bar{V}_{\rm xc}$ of Eq.(\ref{mottbias}) (shown for the same parameters in
panel a) of the same figure) with one crucial difference between
$V_{\rm xc}$ and $\bar{V}_{\rm xc}$: around $I/\gamma \approx 0$ by construction
the xc bias $V_{\rm xc}$ becomes flat in order to satisfy the
FL conditions. However, looking at the step height of $V_{\rm xc}$ of panel c)
we find that it is reduced compared to the step in $\bar{V}_{\rm xc}$. This
latter step is given by the derivative discontinuity $\Delta_{\rm xc}$ and has
the clear physical meaning of the Mott gap, a fundamental physical quantity
of the system which should not be modified by the doping. We can also
understand why our ansatz for $V_{\rm xc}$ leads to the reduction in the step
height: the Kondo prefactor (which multiplies $\bar{V}_{\rm xc}$ in our ansatz)
is just not yet close to unity when the current is close to the
value where the step appears at $I^{\rm stp}/\gamma = 2 (1-n)$. Since by
construction the Kondo prefactor is required to act close to zero current,
this suggests to correct it (for $U>U_c$) in the following way
($j \in \{ 1,2\}$): 
\be
a_{\rm K,j}^{\rm corr.} = \Theta(I^{\rm stp}-I) \, a_{\rm K,j} + \Theta(I-I^{\rm stp})
\hbox{\hspace*{1cm} for $n<1$}
\label{ak_corr_less1}
\ee
and 
\be
a_{\rm K,j}^{\rm corr.} = \Theta(I-I^{\rm stp}) \, a_{\rm K,j} + \Theta(I^{\rm stp}-I)
\hbox{\hspace*{1cm} for $n>1$}
\label{ak_corr_larger1}
\ee
where $\Theta(x)$ is the Heaviside step function. 

In panel d) of Fig.~\ref{fig:qpweight_xcbias} we show the xc bias obtained
with these corrected prefactors. As desired, the height of the step is
now restored to its original value while preserving the FL conditions around
zero current. For $n<1$ ($n>1$) and $I>I^{\rm stp}$ ($I<I^{\rm stp}$) the
new $V_{\rm xc}$ is now identical to the Mott xc bias $\bar{V}_{\rm xc}$. 
Since for both versions of the Kondo prefactor, the corresponding xc biases 
satisfy the FL conditions, around zero current they are
indistinguishable. At closer inspection away from zero current (not shown)
one notes that the xc bias using $a_{\rm K,2}$ approaches $\bar{V}_{\rm xc}$
faster than the one with $a_{\rm K,1}$ which is simply due to the former
approaching unity faster than the latter for large arguments.

\section{Numerical results}

\begin{figure}
  \includegraphics[width=0.9\linewidth]{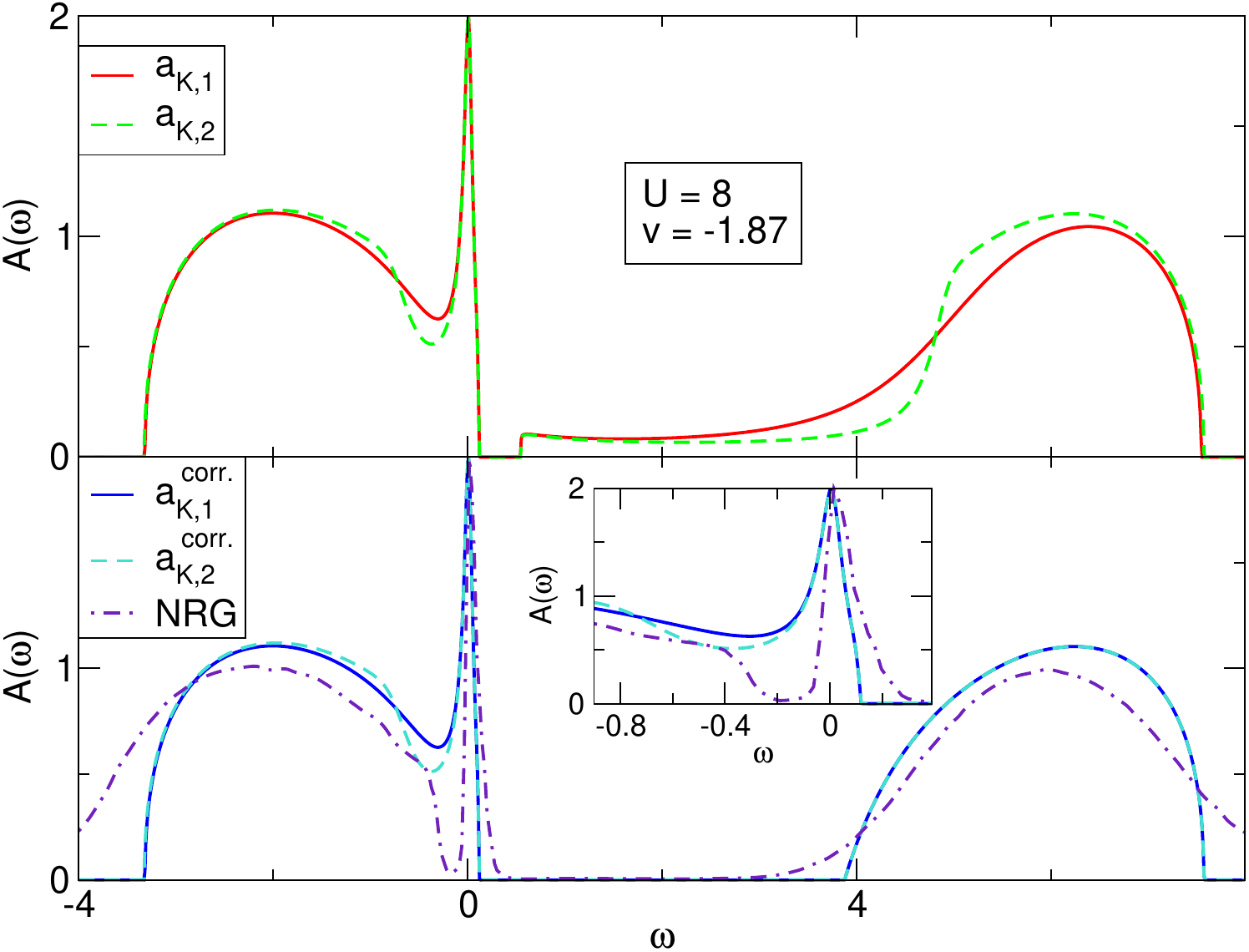}
  \caption{Spectral function of the Hubbard model on the Bethe lattice with
    infinite coordination for $U=8$ and on-site energies $v=-1.87$ (which
    corresponds to a doping of $\delta = 1-n \approx 0,02$). Upper panel:
    i-DFT spectra obtained with the functional of Eq.~(\ref{xcbias-ansatz})
    with the two uncorrected Kondo prefactors $a_{\rm K,1}$ and $a_{\rm K,2}$ 
    (Eqs.~(\ref{akondo1}) and (\ref{akondo2}), respectively). Lower panel:
    i-DFT spectra with the corrected Kondo prefactors of
    Eq.~(\ref{ak_corr_less1}) as well as the NRG spectrum of
    Ref.~\cite{ZitkoHansenPerepelitskyMravljeGeorgesShastry:13}. The inset
    zooms in around the Kondo resonance.}
  \label{fig:spectra_U8_uncut_cut}
\end{figure}

In this Section we show results for the spectral function obtained with
i-DFT using the approximations for the xc bias discussed in the previous
Section. In Fig.~\ref{fig:spectra_U8_uncut_cut} we show i-DFT spectra
for $U=8$ which leads to a derivative discontinuity of 
$\Delta_{\rm xc} \approx 3.42$. For the onsite potential we choose $v=-1.87$
which leads to a doping of $\delta=1-n \approx 0.02$. In the upper panel of
this figure we show spectra obtained with the uncorrected Kondo prefactors
$a_{\rm K,j}$, $j \in \{1,2\}$ (Eqs.~(\ref{akondo1}) and (\ref{akondo2})) while
the lower panel shows the spectra obtained with the corrected Kondo
prefactors of Eq.~(\ref{ak_corr_less1}). In the lower panel we also show the
NRG spectrum of Ref.~\cite{ZitkoHansenPerepelitskyMravljeGeorgesShastry:13}
for comparison. The inset in the lower panel zooms in around the Kondo
resonance. 

We see that for the lower Hubbard band the different Kondo prefactors
give similar results. First of all, the appearance of the Kondo resonance
is captured due to the incorporation of the FL conditions in the xc bias.
The dip in the spectra just to the left of the Kondo resonance is more
pronounced for $a_{\rm K,2}$ (both the uncorrected and corrected versions)
as compared to $a_{\rm K,1}$ and thus more in line with the NRG results while
the width of the Kondo resonance is reasonably well reproduced in i-DFT with
both prefactors (see inset in lower panel).
The big difference between the i-DFT spectra obtained with uncorrected
and corrected Kondo prefactors is in the size of the Mott-Hubbard gap
as well as in the shape of the upper Hubbard band. For the uncorrected
Kondo prefactors the Mott-Hubbard gap (which should be given by
$\Delta_{\rm xc}$) is severely underestimated, a deficiency which is cured
using the corrected prefactors, as intended. Also the upper Hubbard band 
coincides for the two corrected prefactors since its shape in both cases is
determined solely by the Mott xc bias $\bar{V}_{\rm xc}$.

\begin{figure}
  \includegraphics[width=0.9\linewidth]{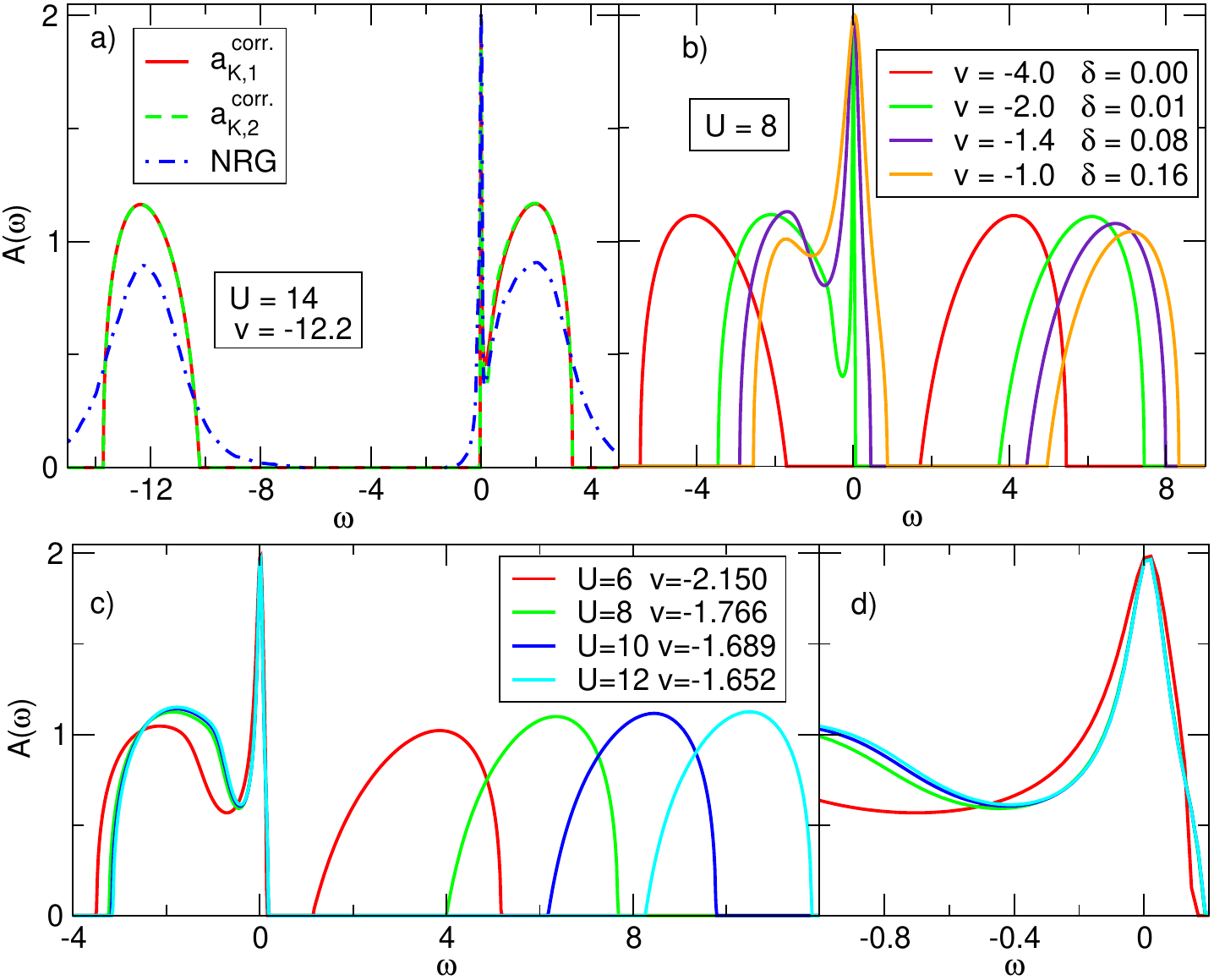}
  \caption{Spectral functions of the Hubbard model on the Bethe lattice with
    infinite coordination. Panel a): i-DFT spectra for $U=14$ and onsite energy
    $v=-12.2$ obtained with the corrected Kondo prefactors of
    Eq.~(\ref{ak_corr_larger1}) compared with the NRG spectrum of
    Ref.~\cite{LeeDelftWeichselbaum:18}. In panels b) to d) only the prefactor
    $a_{\rm K,2}^{\rm corr.}$ is used. Panel b): i-DFT spectral functions
    for $U=8$ for different onsite energies $v$. Panel c): i-DFT spectral
    functions for different values of $U$. The onsite energies were
    chosen such that for all $U$ the doping is the same at $\delta=1-n=0.03$.
    Panel d): same as c) but zoomed in at the Kondo resonance.}
  \label{fig:spectra}
\end{figure}

In Fig.~\ref{fig:spectra} more i-DFT spectral functions are shown.
In panel a) we compare spectral functions for $U=14$ with NRG results
from Ref.~\cite{LeeDelftWeichselbaum:18} with the same overall
agreement as for the case of $U=8$, see discussion of the lower panel of 
Fig.~\ref{fig:spectra_U8_uncut_cut}. In panels b) to d) only the prefactor
$a_{\rm K,2}^{\rm corr.}$ is used. Panel b) shows the evolution of the i-DFT
spectra for $U=8$ when changing the onsite energy. Of course, for
half-filling (doping $\delta = 0$) there is no Kondo resonance which only
appears at finite doping. As expected, with increased doping the resonance
broadens. Panels c) and d) show spectra for different values of $U$ where here
the onsite energies $v$ have been chosen such that the doping $\delta=1-n=0.03$
remains constant. Obviously, the upper Hubbard bands are pushed to higher
energies as $U$ increases because the Mott gap (i.e., $\Delta_{\rm xc}$)
increases from approximately 0.5 for $U=6$ to 7.7 for $U=12$. At the same
time, the quantity $U-\Delta_{\rm xc}$ appears to converge for increasing $U$:
while for $U=6$ we have $U-\Delta_{\rm xc} \approx 0.5$ for $U=6$, $8$, and $10$
the corresponding values for $U-\Delta_{\rm xc}$ are 4.6, 4.4, and 4.3,
respectively. The quantity $U-\Delta_{\rm xc}$ also controls the bandwidth
of the lower and upper Hubbard band. Interestingly, for fixed doping the width
of the Kondo peak in our i-DFT approach also appears to converge as $U$
increases, see panel d). This is probably not a coincidence. For fixed and
small doping the KS potential should be small and the quasiparticle weight
changes essentially linearly with doping and becomes independent of $U$.
Then the only ingredient in $V_{\rm xc}$ which depends on the interaction
is the Mott xc bias $\bar{V}_{\rm xc}$ of Eq.~(\ref{mottbias}).
Assuming that the function $g(x)$ of Eq.~(\ref{gfunc}) is only weakly
dependent on $U$, the remaining dependence of $v_{\rm Hxc}$ (Eq.~(\ref{vhxc}))
and thus $\bar{V}_{\rm xc}$ on the interaction is only through the factor
$U-\Delta_{\rm xc}$. 

\begin{figure}
    \begin{tabular}{cc}
        \includegraphics[width=0.45\linewidth]{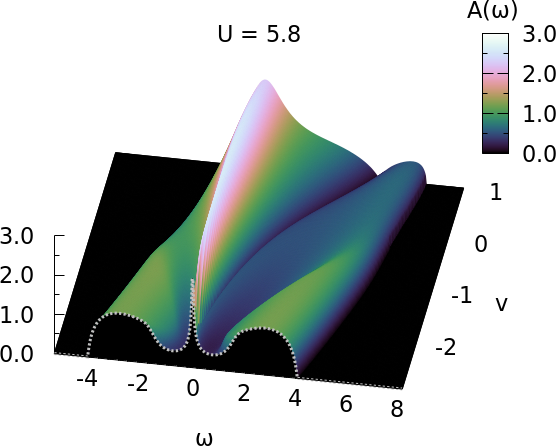} & 
        \includegraphics[width=0.45\linewidth]{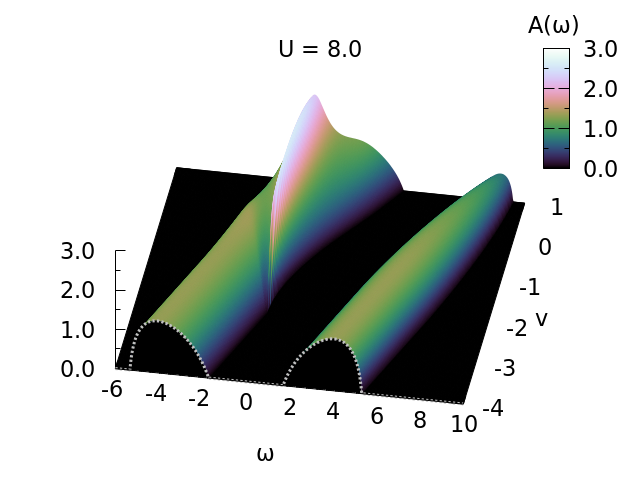} 
    \end{tabular}
    \caption{i-DFT spectral function for the Hubbard model on the
        Bethe lattice as a function of frequency $\omega$ and gate $v$ for
        $U=5.8<U_c$ just in the metallic phase (left) and for $U=8.0>U_c$ deep
        in the Mott phase (right). Here the Kondo prefactor of
        Eq.~(\ref{akondo2}) has been used, along with the corrections proposed 
        in Eqs.~(\ref{ak_corr_less1},\ref{ak_corr_larger1}).
    \label{fig:3d_spectra}
    }
\end{figure}

Fig.~\ref{fig:3d_spectra} shows 3D plots of the spectral function as a function
of $\omega$ and applied gate $v$ for two different values of $U$. 
The left panel shows the spectral function for $U=5.8$, just in the metallic regime
at particle-hole symmetry. At the paticle-hole symmetric point ($v=-U/2=-2.9$)
we thus see the sharp quasiparticle peak at $\omega=0$, together with the two Hubbard side bands
symmetrically placed at $\omega=\pm U/2=\pm2.9$. As $v$ is increased, the
quasiparticle
peak becomes broader, gaining weight and eventually merging with the lower Hubbard band 
as the density approaches zero, $n\to0$. On the other hand, the upper Hubbard band, 
continuously decreases both in weigth and width, eventually disappearing completely as $n\to0$.

The right panel shows the spectral function for $U=8$, deep in the Mott phase. 
Thus at the particle-hole symmetric point $v=-U/2=-4$, there is no quasiparticle 
peak, the spectrum is gapped. As $v$ increases, at first the only effect is the
shifting of both Hubbard bands with the applied gate. But when the critical gate 
for the doping-driven Mott transition, $v_c=\tfrac{U}{2}(\eta_c(U)-1)\sim-2.29$,
is reached, a sharp quasiparticle appears at $\omega=0$. 
Again, as $v$ increase the quasiparticle peak becomes broder and gains weight,
eventually merging with the lower Hubbard band, while width and weight of the upper 
Hubbard decrease, eventually disappearing completely as $n\to0$.

Fig.~\ref{fig:3d_spectra} also shows some limitations of our approximation to the xc bias functional.
In the limit of low occupation, $n\to0$, the spectral function does not
converge to the non-interacting (i.e. the KS) spectral function given by
Eq.~(\ref{sf}). Most likely, this is due to our parametrization of the
quasiparticle weight not approaching unity fast enough in the limit $n\to0$.
Hence $Z^{-2}-1$ in the numerator of Eq.~(\ref{lambda2}) does not vanish
fast enough in the limit $n\to0$ to compensate both for $A_s(0)^4$ in the 
denominator going to zero extremely fast as well as for the 
divergence of $A_s^\prime(0)$ in the same limit.
This prevents $\lambda_2$ from vanishing in this limit, as it should. 
We suspect that the problem can
be fixed by a more careful parametrization of $Z$ in this limit.

\section{Conclusion}

In the present work we have shown how one can capture the doping-driven Mott 
transition within the framework of steady-state density functional theory (i-DFT). 
By construction, standard DFT in principle gives access to the many-body density 
of an interacting many-electron system but requires (an approximation of) the 
Hxc potential as functional of the equilibrium density. A central but rarely used 
tenet of standard DFT is that since the external potential of the many-electron 
system is a functional of the density, in principle {\em all} eigenstates of 
the many-body Hamiltonian (and properties derived from these eigenstates) are 
functionals of the density. Of course, typcially the functional form of these 
functionals is unknown. 

In the so-called ``ideal STM limit'', 
i-DFT gives access to equilibrium many-body spectral functions provided (an 
approximation of) the xc contribution to the bias is known as functional of both 
density and current. Since in this limit, the i-DFT equations for density and 
current are completely decoupled, conceptually one may view the i-DFT access to 
spectral functions as a particular way to write the many-body spectral function 
as a {\em functional} of the equilibrium density, an explicit form 
of which is given by Eq.~(\ref{specfunc_idft}).

In the present work we have constructed an approximate functional for the xc 
bias for interacting electrons on the Bethe lattice with infinite coordination. 
This model plays a central role in DMFT and may be viewed as one of the minimal 
models to study the Mott metal-insulator transition. 
In order to construct an approximate xc bias, we first established 
a connection between the (many-body) quasiparticle weight and the xc bias to 
lowest orders in the current by employing an expression for the Green function 
of Fermi Liquid theory around zero frequency. We used these analytical FL 
conditions to fix two parameters in our ansatz for the xc bias. Due to this 
construction one can expect the i-DFT spectra to reproduce the slope and second 
derivative of many-body spectra around zero frequency, i.e., the location of the 
quasiparticle resonance (if present). Other important ingredients of our functional are 
(i) the famous xc contribution to the derivative discontinuity of DFT which we 
could, for our model system, connect to the plateau of the density as function of 
onsite energies studied with DMFT+NRG in Ref.~\cite{LoganGalpin:16} and (ii) 
the quasiparticle weight as function of the density obtained with NRG 
\cite{Hewson:16} for which we found a simple parametrization.

The i-DFT spectra computed with our approximation for the xc bias reproduce the 
main features of reference spectra of systems with small doping such as the Kondo 
resonance and the Hubbard  sidebands reasonably well. One feature which our 
functional does not capture very well is the pronounced dip between the Kondo 
resonance and the lower (upper) Hubbard band for densities $n<1$ ($n>1$). We have 
tried two forms of the prefactor multiplying the Mott part of the xc bias in our 
ansatz. Compared to the first Kondo prefactor, the second one which approaches 
unity faster for larger arguments leads to somewhat stronger dips than the first 
one. However, even for this prefactor the dips are still not as pronounced as in 
the NRG results. 

In general, our i-DFT approach for the calculation of many-body spectral functions 
is computationally extremely efficient. It requires the solution of (i) a 
standard KS equation to obtain the self-consistent density of the system and (ii) 
for each frequency, i.e., bias, the self-consistent solution of another 
self-consistent equation for the current. This computational efficiency allows 
for rapid multi-parameter scans as, e.g., scans of both onsite energies and 
frequencies. 

Another important result of the present work is the establishment of a theoretical 
connection between the quasiparticle weight, a quantity well-known from many-body 
Green function theory and the xc bias, which is the quintessential i-DFT quantity 
which needs to be approximated in practice. We expect this connection to be 
useful for applications of i-DFT to model systems beyond the one studied here.

Our work also points the way for future research towards an {\em ab initio}
description of Mott insulators and strongly correlated materials in general.
For instance, one could envision mapping the standard DFT electronic structure
of a correlated material onto a few level model which then can be treated
with i-DFT. Possible difficulties along the way could be the proper
extraction of model parameters that avoid the notorious double-counting 
problem of DFT++ approaches~\cite{Karolak:JESRP:2010}, perhaps by making 
use of a different type of mapping between lattice DFT and real-space 
DFT~\cite{SobrinoJacobKurth:23}.
A further challenge could be the construction of a reasonable approximation 
of the quasiparticle weight as a functional of the density.

\section*{Acknowledgements}

\paragraph{Funding information}
D.J. acknowledges funding from the Generalitat Valenciana through the Plan Gen-T of Excellence, 
grant number CIDEXG/2023/7.
S.K. acknowledges financial support from the Basque Government (Eusko Jaurlaritza) 
through the Elkartek program (project CICe2025, grant number KK2025-00054).










\end{document}